# Friction and friction heat of micronscale iron

Le Van Sang[a,**], Akihiko Yano[b], Shuji Fujii[b], Natsuko Sugimura[a,c], Hitoshi Washizu[a,*]

[a]Graduate School of Simulation Studies-University of Hyogo

[b]Mitsubishi Heavy Industries, Ltd.

[c]Faculty of Engineering-Tokyo City University

[*]Email: h@washizu.org

[**]Email: levansang82@gmail.com

The paper investigates friction and friction heat of the micronscale iron under influences of velocity of the slider and temperature of the substrate by using smoothed particle hydrodynamics simulations. In the velocity range of 10 – 100 m/s, change of friction coefficient via velocity well complies with exponent or hyperbolic tangent form and friction coefficient begins to approach a stable value of 0.3 at around a velocity of 50 m/s after a rapidly increasing situation. Friction coefficient steady maintains over the temperature range of 200 – 400 K at each velocity of 10, 50 or 100 m/s. Friction heat is detailed analyzed via sliding time. Change of temperature of the system via sliding time well complies with sigmoidal functions, an exception of that of the particle layer directly causing friction. The layer causing friction has the highest steady temperature and its temperature rise is the largest one. The temperature rise is found to be dependent on increment of the initial temperatures of the substrate and the slider while the increment does not effect on configuration of the sliding time – temperature curves.

## 1. Introduction

Friction and friction heat are always associated with each other in a relative sliding between two objects. Friction has been well-known to be an aspect generating friction heat and in opposite direction friction heat results in friction coefficient. Consisting of both in a study is in need. In general, one should not report friction coefficient of a certain materials pairs without clearly indicating state of contact surface and detailed conditions of study since friction complicatedly depends on rough shapes or smooth ones of contact surface, dry or lubricated sliding conditions, type of lubrication materials, externally applied normal load, coatings, sliding velocity, temperature, so on. This makes wide regions of friction research and interests of this scientific branch. Friction coefficient dependent on sliding velocity has extensively been investigated up to now. Different scenarios have been reported for the velocity-friction curve. In behaviors of increasing sliding velocity over a certain range, friction coefficient exhibited small oscillation around a stead value [1], an approximately linear increase [2,3] or an abrupt growth in the initial stage of the sliding after which occurrence of stable state was established [4,5]. In the case of layered materials, the velocity dependence is more complicated due to the breakage of crystal structure [6,7]. Increase or decrease of friction coefficient via sliding velocity also depended on externally applied normal load [8]. Besides slider velocity, substrate temperature has been found to effect on friction of various materials. Dunckle et al. reported an approximately linear increase of friction coefficient with substrate temperature, $\mu = 0.28$ at $T = 90$ K and $\mu = 0.38$ at $T = 275$ K, in an interfacial friction study of a single-asperity diamond contact on a diamondlike carbon substrate [9]. However, Pearson et al. found that the steady state friction coefficient monotonically dropped with increasing temperature, from 0.78 at 24 ºC to a minimum of 0.46 at 450 ºC [10]. Thus, friction coefficient should be reported by associating with clear details of velocity, temperature and a given contact materials pairs. The present work investigates influences of velocity of the slider and temperature of the substrate on friction coefficient of the coarse-grained microscale iron system in some simulation conditions.

It is well-known that increase of friction heat causes significant influences on contact surface such as structural transformation, oxidation, wear, friction and melting. Due to its major influences on tribological phenomena, friction heat was commonly investigated in the past. One can make economic and stable tribological designs by having good understandings in friction and friction heat of contact materials pairs. Most theoretical, experimental and simulation studies stated that temperature of contact surface abruptly increased in the initial stage of the sliding after which it stably maintained via sliding time [11-15]. Due to some difficulties facing in experimental approach for determination of contact temperature [16,17], using simulation methods to do this work can be a noted consideration. Li et al. used molecular dynamics simulations to observe generated heat during friction sliding of the Ni-Al, Al-Ni and NiAl-NiAl pairs [18]. Results of finite element simulation and experiment are close to each other in sliding time dependence of contact temperature for the polyimide-stainless steel [19] and the WC-6Co pin-steel disc AISI 1045 pair [11]. However, most previous studies have not ever carried out a rule (or a function) of the sliding time – friction coefficient/heat curve. Distribution of friction heat and heat transfer on the system during the sliding are carefully surveyed in our present work.

## 2. Calculations and simulations

The $N = 21800$ particles coarse-grained iron system comprises the substrate of 75.240×16.245×7.695 ($\mu m^3$) and the slider of 16.425×16.245×7.695 ($\mu m^3$) (Fig. 1). The particle system is obtained from coarse-graining of the body-centered-cubic atomic iron system as done in our previous work [20]. In the present work the lattice constant of the particle system is $a_{CG} = 0.855$ μm and the mass of particle is $m = 5.01 \times 10^{-6}$ μg, corresponding to the coarse-graining of $N_l = 3000$ seen in our previous work [20]. In Fig. 1 the system is divided into five layers characterized by different colored particles in the z-direction, .i.e. which are named L1, L2, L3, L4 and L5 from bottom to top, respectively. Along with the x-sliding direction the L3 is divided into four regions of the equal length of 16.425 μm, i.e. which are named G1, G2, G3 and G4 from left to right, respectively. The same as done for the L3, the L2 is also divided into four regions named G5, G6, G7 and G8. These divisions make convenient discussions of temperature during the sliding.

Smoothed particle hydrodynamics method is used in the present work. In this approach the density $\rho_i$, velocity $v_i^\alpha$ and internal energy $u_i$ of the $i$th particle are presented by the first derivative equations of time

$$\frac{d\rho_i}{dt} = \sum_{j=1}^{N} m_j \left( \vec{v}_j - \vec{v}_i \right) \vec{\nabla}_i W \left( \vec{r}_{ij}, h_{ij} \right),$$ (1)

$$\frac{dv_i^\alpha}{dt} = \sum_{j=1}^{N} m_j \left( \frac{\sigma_i^{\alpha\beta}}{\rho_i^2} + \frac{\sigma_j^{\alpha\beta}}{\rho_j^2} + \Pi_{ij} \right) \nabla_i^\beta W \left( \vec{r}_{ij}, h_{ij} \right),$$ (2)

$$\frac{du_i}{dt} = \frac{1}{2} \sum_{j=1}^{N} m_j \left( \frac{\sigma_i^{\alpha\beta}}{\rho_i^2} + \frac{\sigma_j^{\alpha\beta}}{\rho_j^2} + \Pi_{ij} \right) \left( v_j^\alpha - v_i^\alpha \right) \nabla_i^\beta W \left( \vec{r}_{ij}, h_{ij} \right) + \sum_{j=1}^{N} \frac{m_j}{\rho_i \rho_j} \frac{4\kappa_i \kappa_j}{\kappa_i + \kappa_j} \left( T_i - T_j \right) \frac{\vec{r}_{ij} \vec{\nabla}_i W \left( \vec{r}_{ij}, h_{ij} \right)}{r_{ij}^2 + \epsilon h_{ij}^2}$$ (3)

where $\alpha, \beta \equiv x, y, z$; $m$, $\kappa$ and $T$ are mass, thermal conductivity and temperature of particle, respectively; $\vec{r}_{ij} = \vec{r}_i - \vec{r}_j$ is relative position vector between the particles $i$ and $j$; $h_{ij}$, $\sigma^{\alpha\beta}$ and $\Pi_{ij}$ correspond to smoothed length, stress tensor and artificial viscosity function, which are considered the same those in our previous work [20]; and $W \left( \vec{r}_{ij}, h_{ij} \right) = 1/(\pi^{3/2} h_{ij}^3) \left( 5/2 - r_{ij}^2 / h_{ij}^2 \right) \exp(-r_{ij}^2 / h_{ij}^2)$ is kernel function. We consider adding the dissipation force on each particle of the system to compensate energy dissipation caused by friction during the sliding

$$F_{dis,\,i} = \begin{cases} -m_i \gamma_{dis} \left( v_i^x - V_{dis} \right) & \text{the x-direction} \\ -m_i \gamma_{dis} v_i^y & \text{the y-direction} \\ -m_i \gamma_{dis} v_i^z & \text{the z-direction} \end{cases}$$ (4)

where $\gamma_{dis}$ is a parameter of the model and $V_{dis} = 0$ for particles of the substrate and $V_{dis} = V$, which is a constant sliding velocity of the slider in the x-direction, for particles of the slider. By using the Prandtl-Tomlinson model, we also add a spring force on each particle of the slider

$$F_{spr,i} = \begin{cases} K\left(x_{0,i} + Vt - x_i\right) & \text{the x-direction} \\ K\left(y_{0,i} - y_i\right) & \text{the y-direction} \\ K\left(z_{0,i} - z_i\right) & \text{the z-direction} \end{cases} \tag{5}$$

where $K$ is a spring constant, $t$ is sliding time, $x_{0,i}$, $y_{0,i}$ and $z_{0,i}$ are the equilibrium/initial coordinates of the $i$th particle in the x, y and z-directions, respectively. Interaction between the slider and the substrate is presented by interaction between particles of the L4 and particles of the L3. Two particles, one of each layer, interact with each other via the following spring force

$$\vec{F}_{int,ij} = \begin{cases} -k_\alpha\left(r - h\right)\dfrac{\vec{r}_{ij}}{r} & 0 < r < h \\ 0 & r > h \end{cases} \tag{6}$$

where $k_\alpha$ is an anisotropic spring constant. The friction force $F_{fri}$, the normal force $F_{nor}$ and the friction coefficient $\mu_{cof}$ are defined as follows

$$F_{fri} = \sum_{i=1}^{N_f} \left(F_{spr,i}^x + F_{int,ij}^x\right), \tag{7}$$

$$F_{nor} = \sum_{i=1}^{N_f} \left(F_{spr,i}^z + F_{int,ij}^z\right), \tag{8}$$

$$\mu_{cof} = \frac{F_{fri}}{F_{nor}} \tag{9}$$

where $N_f$ is the number of the particles of the L4, $F^x$ and $F^z$ are the force components in the x and z-directions, respectively. Temperature of the $i$th particle is estimated by the formula

$$\frac{dT_i}{dt} = \frac{u_i + u_{fri,i}}{C_i^P} \tag{10}$$

where $C_i^P$ is heat capacity of the particle and $u_{fri,i}$ is friction energy, noticing that $u_{fri} = 0$ for the particle different from the friction interacting particles (Eq. (6)).

The initial distance between the slider and the substrate is equal to $a_{CG} = 0.855$ µm. Particles of the L1 are fixed during the simulations. The slider slides a segment of 55.575 µm along with the x-direction in each simulation. The parameters (several of which are contained in $h_{ij}$, $\sigma^{\alpha\beta}$ or $\Pi_{ij}$) are used in each simulation: $\mu = 52.5$ GPa [21], $\eta = 1.2$, $A = B = 0.1$, $h_{cr} = 1.4$, $\epsilon = 0.01$, $\gamma_{dis} = 10^3$ 1/s, $\rho = 7.86$ g/cm$^3$, external pressure of 0.1 MPa, a time step of $dt = 285$ ps, and $K = 0.051$ nN/nm for a sliding velocity of 100 m/s as done in our previous work [20] and referenced from [22]. We consider fixing the term contributed by sliding velocity in the Prandtl-Tomlinson model (Eq. (6)) when we employ with different sliding velocities, meaning that $KV$ is setup to a constant value. The anisotropy of the spring force in Eq. (7) is presented by $k_z = k_y = 0.1k_x$ with $k_x = 0.2K$ in each simulation. This anisotropy is the best behavior for investigating sliding friction of coarse-grained micronsclae iron [20]. The data of the thermal conductivity and the heat capacity of iron are taken from works of Powell et al. [23] and Desai [24], respectively. When simulation temperature of particle is closer to a referred temperature of $\kappa$ and $C^P$ in the data, these values ($\kappa$ and $C^P$) are then setup to the particle at this simulation temperature. We modify the FDPS open source developed by Iwasawa et al. [25] to create our simulation program.

## 3. Results and discussions

In order to consider influences of temperature on friction of the system in the smoothed particle hydrodynamics method, the three different simulations at $V = 50$ m/s are carried out as follows: the internal energy does not include the temperature term of Eq. (3) (the second term of r.h.s of Eq. (3)); the internal energy is taken from Eq. (3), the initial temperature of the slider is $T = 300$ K and that of the substrate is $T = 302$ K; or $T = 305$ K, .i.e. which are named Case1, Case2 and Case3, respectively. The system shows a regularly stick-slip motion during the sliding in all the cases (Fig. 2). Distance between the nearest peaks of each curve is exactly equal to the lattice constant of the system. These scenarios are similar to those reported in our previous work that excluded temperature dependence of the internal energy [20]. Notice that the stick-slip sliding can still occur at high velocities, 100 m/s for a micron iron-iron system [20] and 30 or 60 m/s for an atomistic-scale Al-Al system [26]. Each curve matches closely to each other in its oscillation amplitude and average value, approximating 0.3 seen in the inserted figure, which is considered as friction coefficient of the system in the below discussions. This demonstrates that in the smoothed particle hydrodynamics method a small increment of initial temperatures of a slider and a substrate does not result in friction coefficient of a system. This state can also be explained by molecular dynamics approach in which velocity and temperature of particle comply with the rule $V = \sqrt{3k_B T / m}$ where $m$ is mass of particle and $k_B$ is the Boltzmann constant. In our behaviors the initial temperatures of the substrate are low, their difference (3 K) is small and the mass of particle is large; therefore, under a molecular dynamics view velocity of particle shows a little difference between the behaviors, leading that structure of the system or its contact surface is near same in the behaviors. This leads to coincidence of friction coefficient as observed. The obtained friction coefficient of iron, $\mu = 0.3$, is consistent with that reported in previously experimental studies [27,28]. Fig. 3 shows the sliding velocity dependence of friction coefficient for Case2. Unlikely the Coulomb model that stated that friction coefficient is not dependent on sliding velocity, our result shows that friction coefficient is an increasing function of sliding velocity and its velocity-evolution is consistent with that in approximate Coulomb models such as saturation $\mu = \mu_0 \mathrm{sat}\left(\vartheta V\right)$ [29], arctangent $\mu = \mu_0 \mathrm{atan}(\vartheta V)2/\pi$ [30] and hyperbolic tangent $\mu = \mu_0 \mathrm{tanh}\left(\vartheta V\right)$ [31] where $\mu_0$ is maximum or stable value of $\mu$ and $\vartheta$ is a parameter. Our simulation data has been fitted with exponent and hyperbolic tangent functions, as done by Sideris et al. [32] and Marklund et al. [33]. Both show the good fits. It is also easy to see in the exponent fit that friction coefficient of a microscale iron system starts to approach a stead value of ~ 0.3 at around a sliding velocity of 50 m/s. The state in which friction coefficient goes to reach a stead value at a certain sliding velocity has been found theoretically, numerically and experimentally in contact of a steel–steel couple [4], an elastomer-a rigid rough indenter [5] and disk-type wet clutches [33]. The hyperbolic tangent fit, which gives a fitted function $\mu(V) = 0.1492\tanh(0.0781V) + 0.0398V^{0.1} + 0.0931$, can be better than the exponent one (Fig. 3). This fit also yielded an excellent coincidence between a fitted line and an experimental data for friction coefficient of disk-type wet clutches [33]. It can easily be seen coincidence of the velocity – friction curves among the three considered cases in the inserted one in Fig. 3(c).

Fig. 4 shows the sliding time-evolution of temperature of the layers at $V = 50$ m/s for Case2 (Fig. 4(a)) and Case3 (Fig. 4(b)). The inserted figures are the continuous plots of the main ones via the sliding time. Temperature of each layer is average temperature per its particle estimated by $T_{layer} = 1 / \mathrm{N}_{layer} \sum_{i \in layer}^{\mathrm{N}_{layer}} T_i$ where $\mathrm{N}_{layer}$ is the particle number of the layer. Temperature abruptly increases in the initial stage of the sliding after which it steady maintains during the remaining time in all the layers. This scenario is in accordance with simulation or experimental observations for the WC-6Co pin versus the steel disc AISI 1045 [11], the alumina-steel dry sliding contact [12] and the ultra-high molecular weight polyethylene against CoCrMo alloy or sapphire glass [13]. The strong growth of temperature in the initial stage is mainly contributed by friction heat, not significantly created from an increment of initial temperatures of a slider and a substrate. Our results also indicate that temperature rise of the slider is much faster than that of the substrate and stead temperatures of the contact layers are higher than those of the others. The layer causing friction always has the highest temperature during the sliding. One other notice there is a very slight difference between the stead temperatures of the L3 and L5 because a segment of the L3 is always in contact and the L5 directly takes heat from the L4. In spite of coincidence of friction coefficient between Case2 and Case3, the stead temperature differs from each behavior (Fig. 4(a) and (b)). Subtraction of the stead temperatures of corresponding layer between Case3 and Case2 is close to 3 K being exactly equal to increment of the initial temperatures of the substrate between the two behaviors. Therefore, in addition to heat generation of friction by making a temperature rise of the layers, there is also heat transfer between the slider and the substrate or between the layers. The layer causing friction exhibits the highest temperature rises of about 12.7 and 15.7 K corresponding to Case2 and Case3, while those of the L2 are about 2.7 and 5.7 K. It is also found in Fig. 4 that the time-evolution of temperature complies with sigmoidal functions for all the layers, exception of the L4. Fig. 5 shows the time-evolution of temperature of the regions at $V = 50$ m/s for Case2 (Fig. 5(a)) and Case3 (Fig. 5(b)). Temperature of each region grows as the slider goes to it, reaches the highest one as the slider completely lies on it, and drops as the slider starts to leave out it. Temperature of one region begins to strongly grow as the slider touches it. At contacting point, temperature decreases via the substrate depth proved by comparison of temperatures of the G1 and G5, G2 and G6, G3 and G7, and G4 and G8. This is because that friction heat forms at only a very thin layer of contact surface [34] and heat is transferred from the contact into the bulk. Temperature of a substrate decreasing with its depth has been well-known in studies of sliding friction [19,35]. When the slider just goes out the contact, thermal equality is fast established via depth due to a small increment of temperatures via depth, thermal dissipation into the bulk and heat transfer. Thermal equality at the interface can be clearly seen from approximation of temperatures of the peak of each region (the G1, G2, G3 or G4) (Fig. 5) and the stead temperature of the L4 (Fig. 4). Temperature regularly varies in sliding direction during the sliding.

Fig. 6 shows the temperature dependence of friction coefficient. The data in Fig. 6 are carried out in simulation conditions as follows: the initial temperature of the slider is 300 K, the temperature of the L2 is fixed at each given temperature (from 200 K to 400 K) during the sliding in each simulation, and the initial temperature of the L3 is equal to that of the L2. This behavior is because that the L3 is effected from friction heat which can be a major factor effecting friction coefficient, particularly for frictional materials with low melting point and high thermal conductivity [36]. Our results indicate that at each velocity of 10, 50 or 100 m/s friction coefficient has near no change as temperature of the L2 increases from 200 to 400 K. Some studies also reported the same this state for various materials. Experiments of Perfilyev et al. for friction of CrV(x)N coatings at four vanadium contents of x= 0, 12, 27 and 35% exhibited that an increase in the content of vanadium leaded to a decrease in friction coefficient at temperature of 25 or 500 °C, whereas in the same the content friction coefficient showed a little difference between these two temperatures [37]. The contact surfaces or friction coefficients can be very slight effected by temperature in our and Perfilyev et al.'s studies since the considered temperatures are further below the temperatures of the melting or the pre-melting of iron and chromium nitride. This is in a good agreement with Dias et al.'s report that stated that friction coefficient remained a stead value via temperature prior to the pre-melting temperature of the contact surface at which it abruptly grows [38].

## 4. Conclusions

In the present simulation work of friction and friction heat of coarse-grained micronscale iron, we find that changes of friction coefficient via sliding velocity and temperature of the substrate via sliding time well comply with several basic functions as proved the above. Friction coefficient steady maintains in the substrate temperature interval of 200 – 400 K which is further below the pre-melting point of iron, and begins to approach the stable value of 0.3 at a slider velocity of 50 m/s. A small increment of the initial temperatures of the substrate and the slider does not effect on configuration of the sliding time-temperature curves whereas it results in the temperature rise of the layers. Temperature shows the highest one at the layer causing friction and decreases from the contact surface into the bulk of the substrate or the slider. Thermal equality is found at the interface.

**Figure captions:**

**Fig. 1.** The coarse-grained micronscale iron system of 21800 particles.

**Fig. 2.** Evolution of friction coefficient via sliding time, (a) the temperature term is excluded in the equation of the internal energy (Eq. (3)), (b) the temperature term is included and the initial temperature of the slider is 300 K and of the substrate is 302 K, and (c) the temperature term is included and the initial temperature of the slider is 300 K and of the substrate is 305 K.

**Fig. 3.** Sliding velocity dependence of friction coefficient (CoF) for Case2 and its fitted lines. The inserted one is plotted for the three cases, Case1: up triangle, Case2: circle and Case3: down triangle.

**Fig. 4.** Evolution of temperature of the layers via sliding time for (a) Case2 and (b) Case3.

**Fig. 5.** Evolution of temperature of the regions of the substrate via sliding time for (a) Case2 and (b) Case3.

**Fig. 6.** Friction coefficient dependent on temperature of the layer L2.

**Fig. 1.**

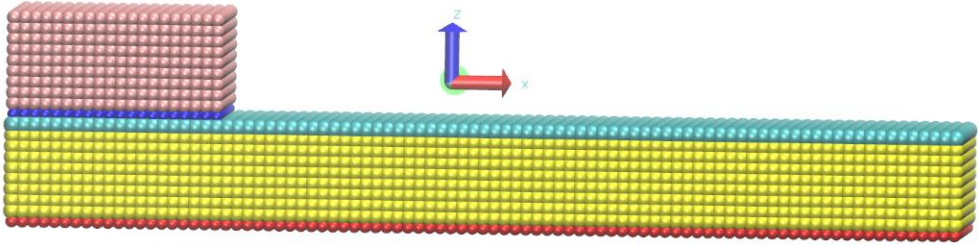

**Fig. 2.**

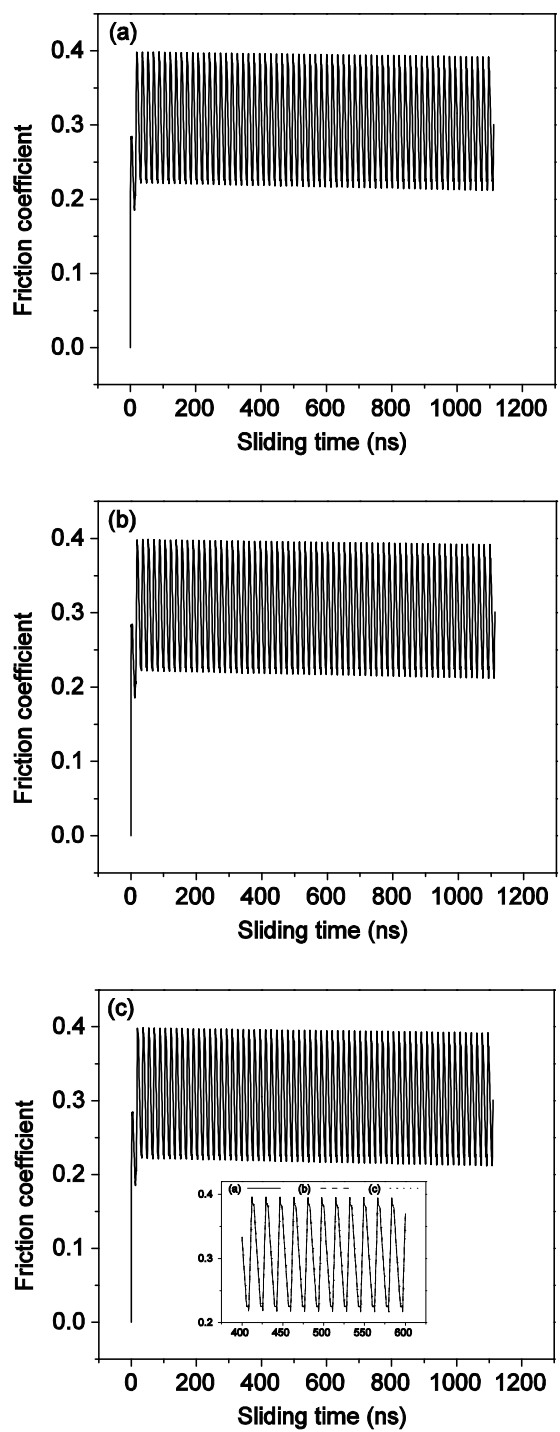

Fig. 3.

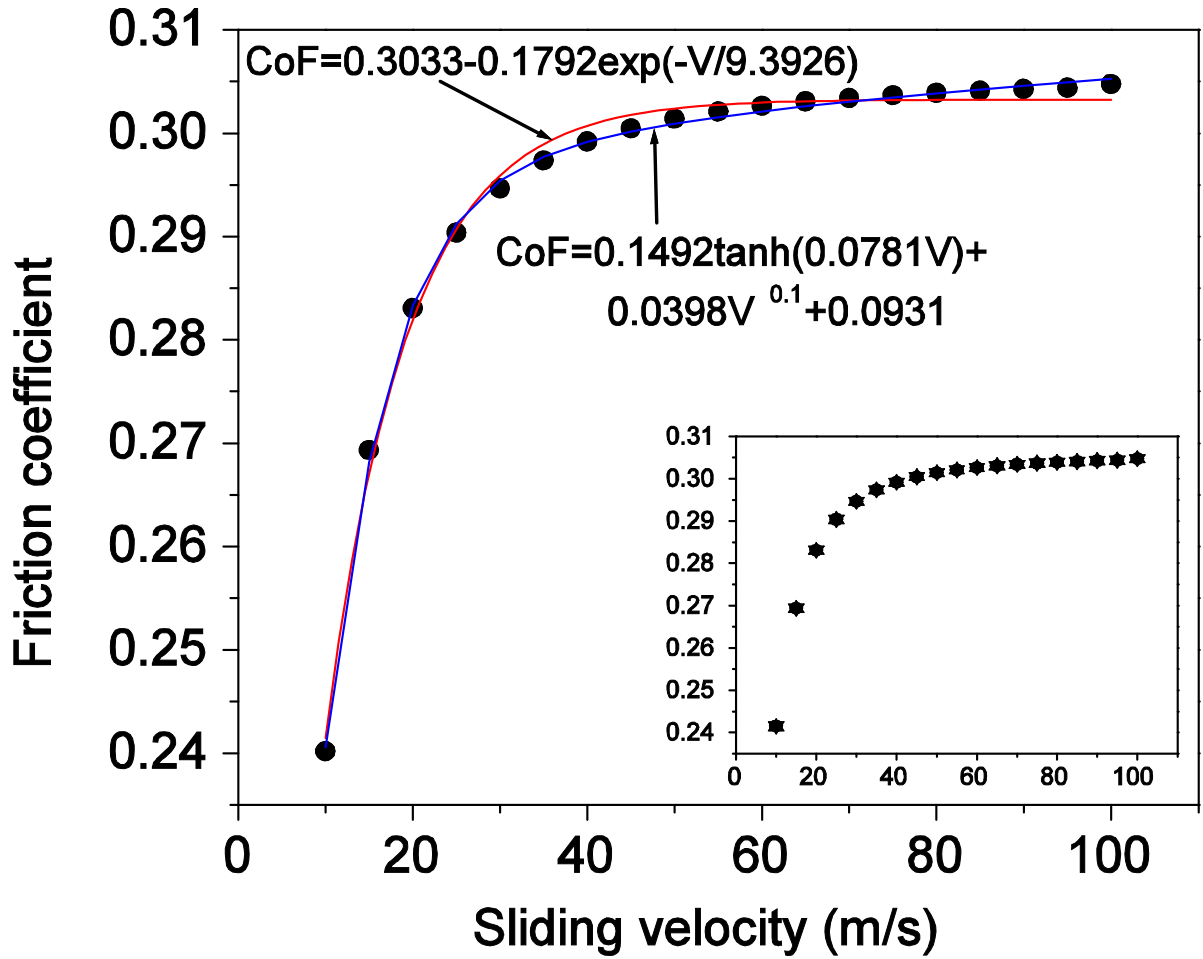

**Fig. 4.**

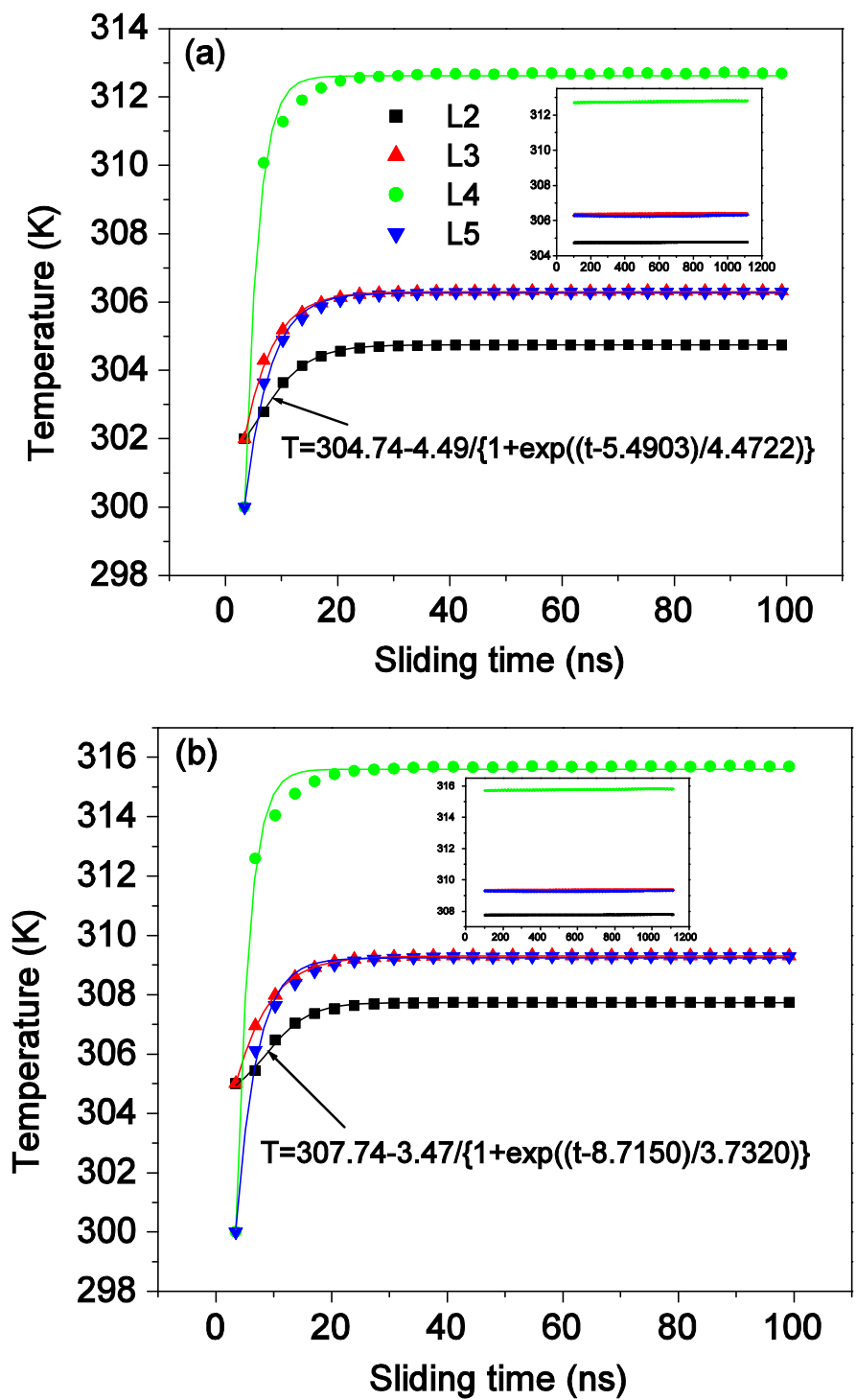

**Fig. 5.**

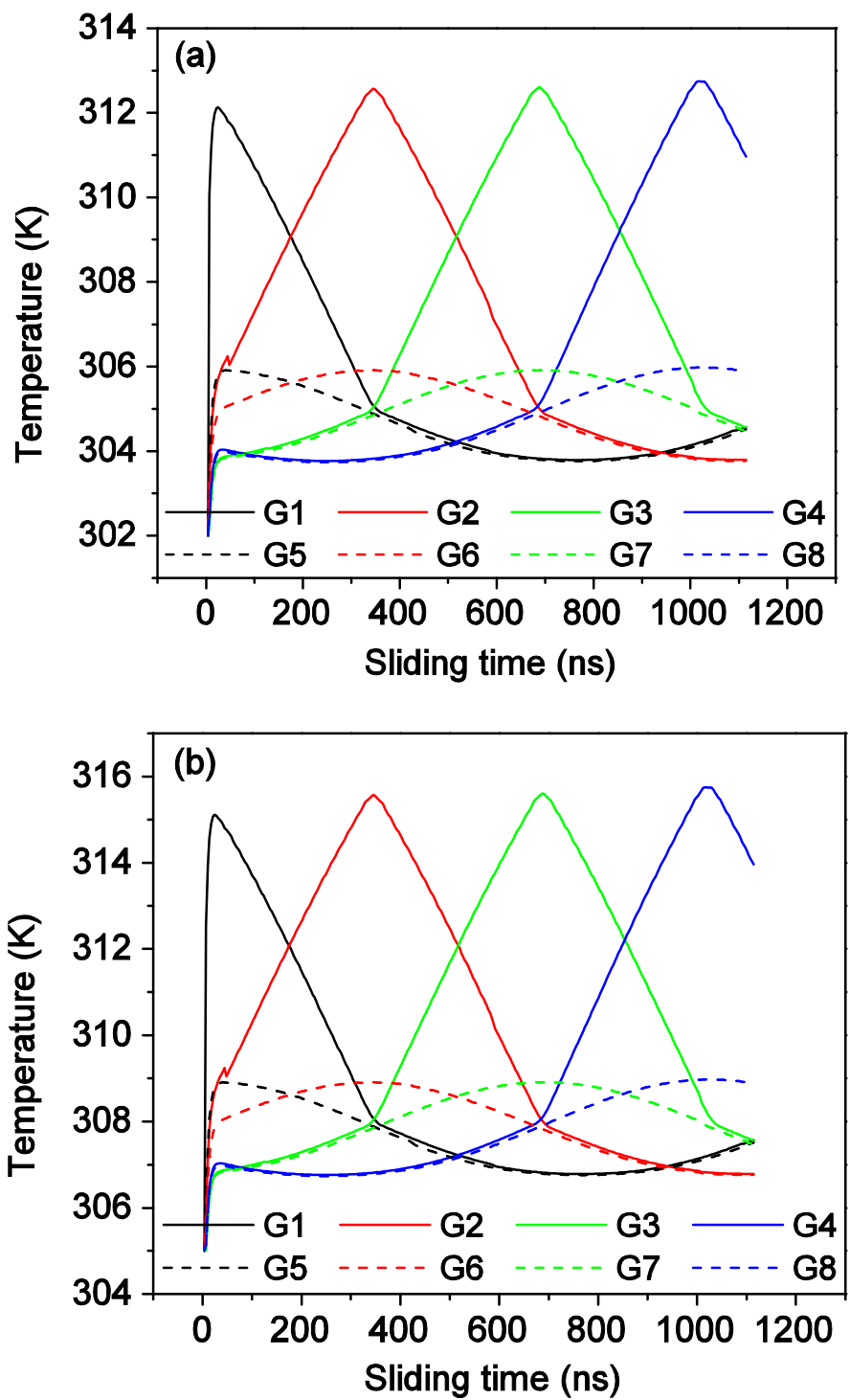

Fig. 6.

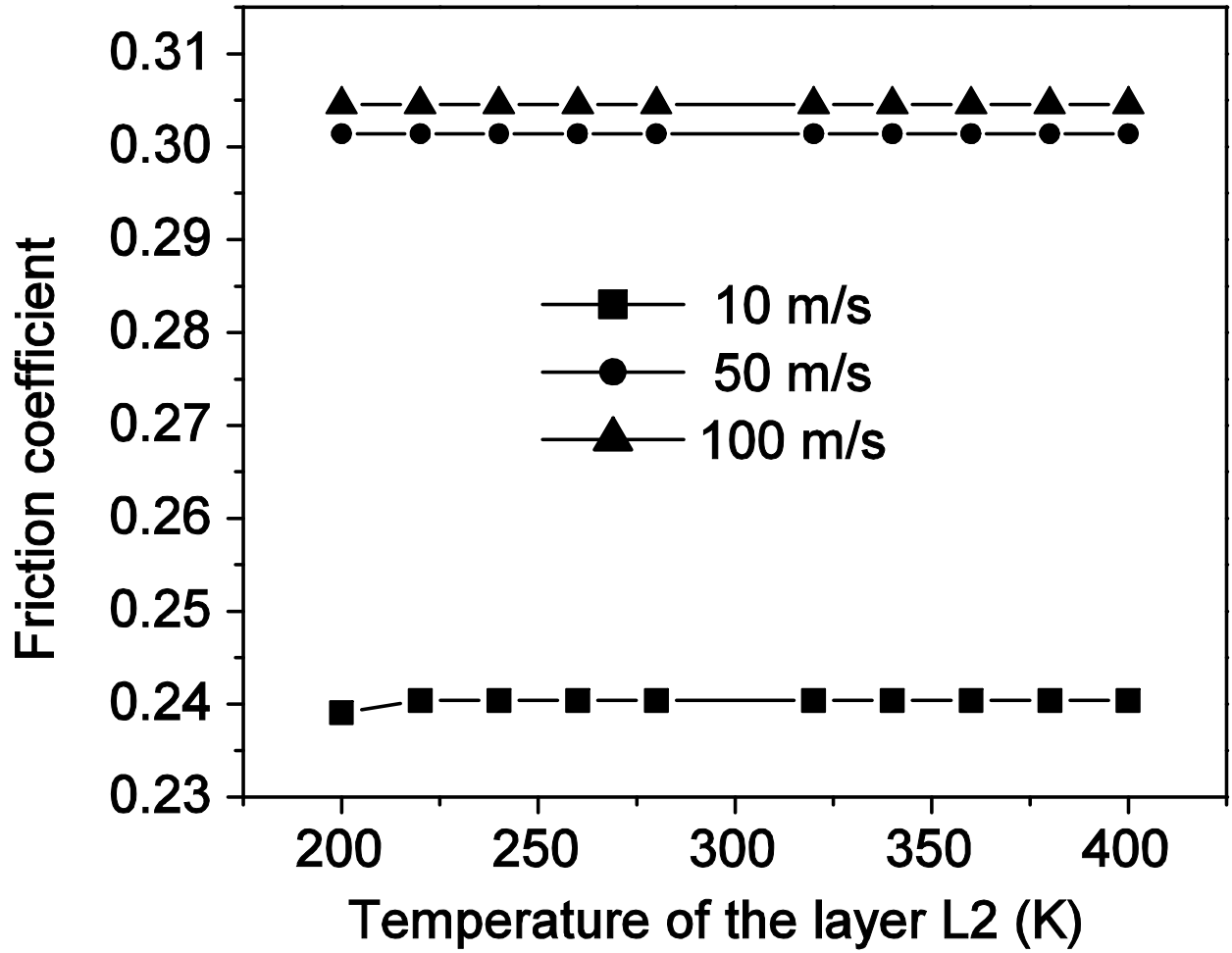